\documentclass[final,3p,times,12pt,]{elsarticle}
\usepackage{amssymb}

\usepackage{graphicx}
\usepackage{subfigure}
\journal{Astroparticle Physics}

\begin{document}
\begin{frontmatter}

\title{ Unexpected enhancement in secondary cosmic ray flux during the total lunar eclipse of December 10, 2011}

\author[1]{Anil Raghav\corref{cor2}\fnref{fn1}}
\ead{raghavanil1984@gmail.com /anil.raghav@physics.mu.ac.in}
\cortext[cor2]{Principal corresponding author}
\fntext[fn1]{ Present address: Department of Physics, University of Mumbai, Vidyanagari, Santacruz (E), Mumbai-400098, }
\author[2]{Ankush Bhaskar}
\author[2] {Virendra Yadav}
\author[1]{Nitinkumar Bijewar}
\author[1]{Chintamani Pai}
\author[1]{Ashish Koli}
\author[1]{Nilam Navale}
\author[1]{Gurinderpal Singh}
\author[1]{Nitin Dubey}
\author[1]{Sushant Pawar}
\author[1]{Pradnya Parab}
\author[1]{Gandhali Narvankar}
\author[1]{Vaibhav Rawoot}
\author[1]{Vikas Rawat}
\author[1]{Satish Borse}
\author[1]{Nagnath Garad}
\author[1]{Carl Rozario}
\author[1]{Nitin Kaushal}
\author[1]{Shailendrakumar Tiwari}
\author[1]{M. R. Press}

\address[1]{Department of Physics, University of Mumbai,Vidyanagari, Santacruz (E), Mumbai-400098, India}
\address[2]{India Indian Institute of Geomagnetism, Kalamboli Highway, New Panvel, Navi Mumbai- 410218, India.}

\begin{abstract}

{Temporal variation of secondary cosmic rays (SCR) flux was measured during the total lunar eclipse on December 10, 2011 and the subsequent full moon on January 8, 2012. The measurements were done at Department of Physics, University of Mumbai, Mumbai (Geomagnetic latitude: 10.6$^\circ$ N), India using NaI (Tl) scintillation detector by keeping energy threshold of 200 KeV. The SCR flux showed approximately 8.1 $\%$ enhancement during the lunar eclipse as compared to the average of pre- and post-eclipse periods. Weather parameters (temperature and relative humidity) were continuously monitored and their correlations with temporal variation in SCR flux were examined. The influences of geomagnetic field, interplanetary parameters and tidal effect on SCR flux were considered.  Qualitative analysis of SCR flux variation indicates that the known factors affecting SCR flux fail to explain observed enhancement during the eclipse. This enhancement during lunar eclipse and widely reported decrease during solar eclipses may unravel hitherto unnoticed factors modulating SCR flux.}

\end{abstract}


\end{frontmatter}

% \linenumbers
%% main text
\section{Introduction}
\label{1}
{
Our planet Earth is being constantly bombarded by high energy particles from the Sun and Galactic Cosmic Rays (GCR). Though the Earth\textquoteright s magnetic field provides a protective shield sustaining the life, still high energy GCR manage to reach the Earth and contribute to Secondary Cosmic Rays (SCR) flux. SCR flux variations have been extensively studied for Solar Cycles, 27 days' cycle, diurnal variations, Coronal Mass Ejections (CMEs) and solar eclipses \cite{Dorman09,Cecchini09}.  SCR flux is known to vary with factors such as local weather parameters (temperature, pressure and humidity), geomagnetic variations, interplanetary parameters and tidal effects. Since 1995, SCR flux variations during solar eclipses have attracted attention and a typical decrease in SCR flux has been reported \cite{Bhattacharyya97,Kandemir00,Antonova07,Bhaskar11,Bhattacharya10,Nayak10}. In such solar eclipse studies, researchers have attempted to correlate weather parameters and geomagnetic variations with SCR flux to understand the underlying mechanism. Of these, the local weather parameters have been thought of as a major factor for the observed decrease as there is a rapid change in local weather parameters during a solar eclipse. Still the complete physical mechanism of the observed decrease remains unraveled. On the other hand, during a lunar eclipse no variation in local weather parameters is expected. Hence one cannot expect any change in SCR flux variation during a lunar eclipse. May be due to this reason less attention has been given to SCR flux variation studies during lunar eclipses. More over the unique geometrical alignment of the Sun, the Earth and the Moon during an eclipse and effective tidal forces may be responsible for the observed decrease during solar eclipses. To investigate the possibilities of these two effects, we carried out SCR flux measurements during the total  lunar eclipse on December 10, 2011 and the subsequent full moon (control day) on January 8, 2012 at Department of Physics, University of Mumbai, Mumbai (Geomagnetic latitude: 10.6$^\circ$ N), India. Subsequent full moon day was purposely chosen as control day given its similar geometry during lunar eclipse.

\begin{figure}[htp]
\centering
\includegraphics[angle=0,width=120mm]{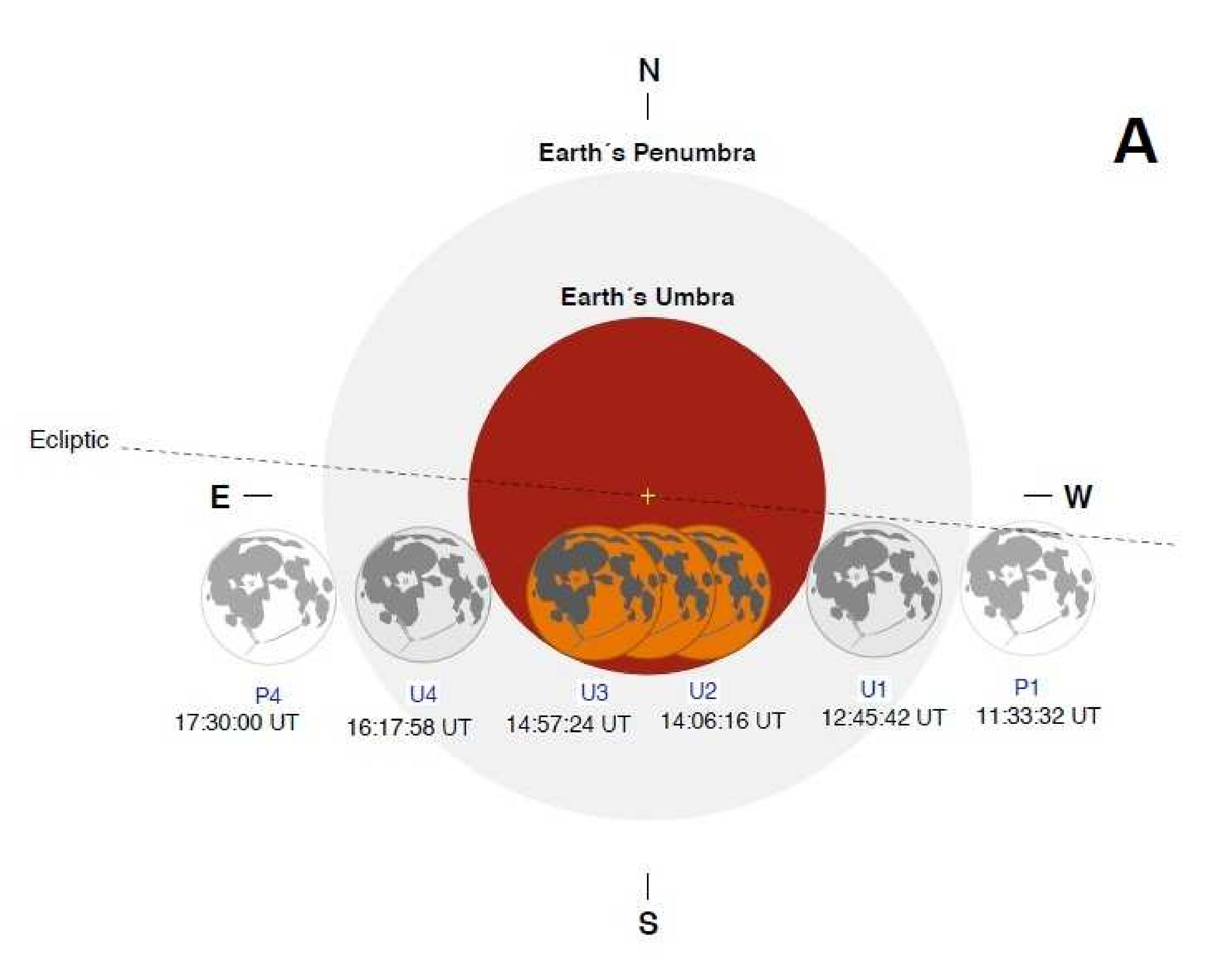}
\caption{Geometrical and Temporal parameters associated with the lunar eclipse where P1,U1,U2,U3,U4 and P4 are the lunar contact timings with the Earth\textquoteright s Penumbra (P) and Umbra (U)(Eclipse map courtesy of Fred Espenak - NASA/Goddard Space Flight Center. See http://eclipse.gsfc.nasa.gov/eclipse.html for more information on solar and lunar eclipses.)}
\end{figure}

\begin{figure}[htp]
\centering
\includegraphics[angle=0,width=120mm]{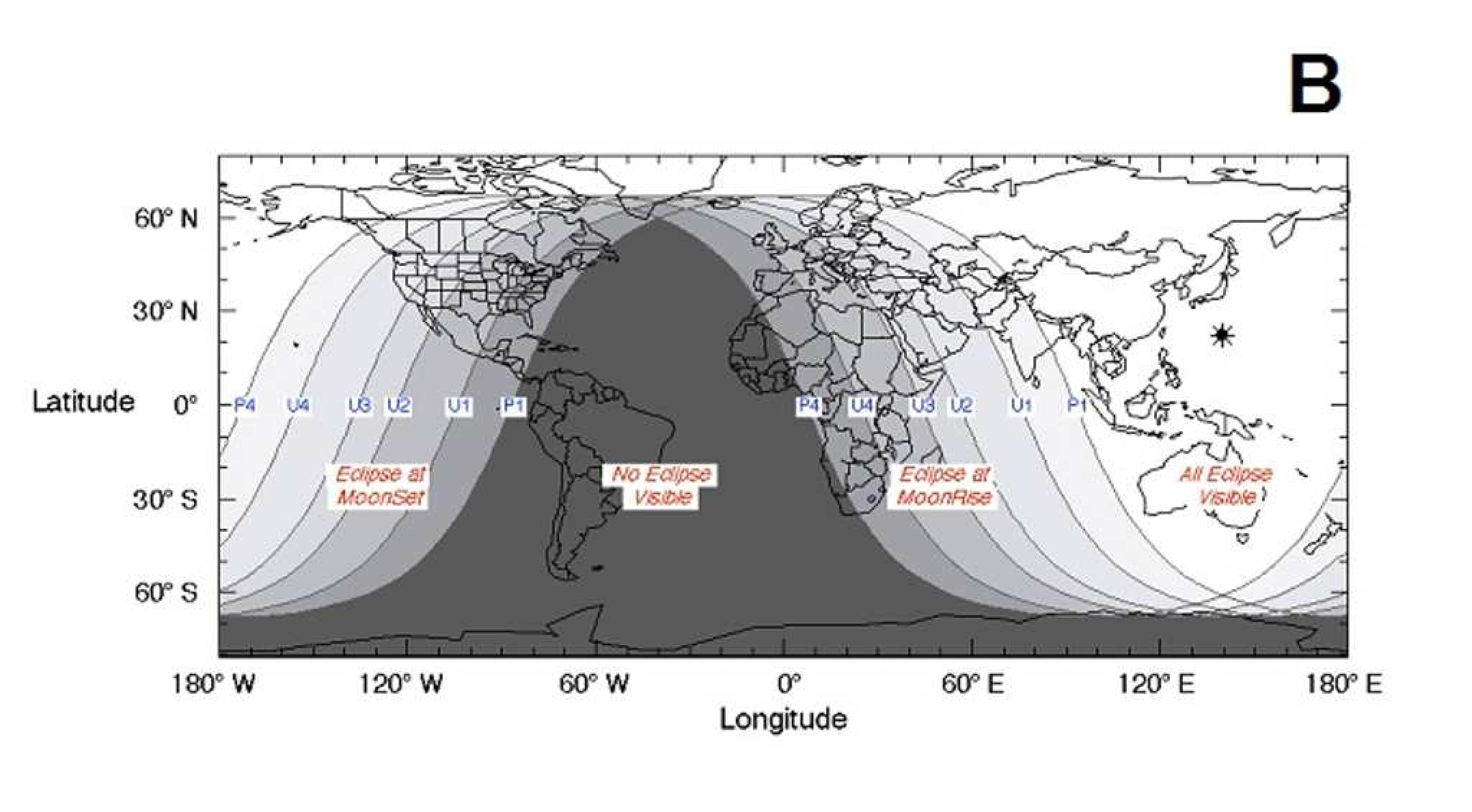}
\caption{World map of the eclipse visibility.(Eclipse map courtesy of Fred Espenak - NASA/Goddard Space Flight Center. For more information on solar and lunar eclipses, see http://eclipse.gsfc.nasa.gov/eclipse.html.)}
\end{figure}

%\caption{Total lunar eclipse on December 10, 2011. (Eclipse map courtesy of Fred Espenak - NASA/Goddard Space Flight Center. For more information on solar and lunar eclipses,see http://eclipse.gsfc.nasa.gov/eclipse.html.  )}

During this eclipse, the Moon\textquoteright s orbital trajectory took it through the southern half of the Earth\textquoteright s umbra. The total duration of the eclipse was about 5 hr 56 min. Although the eclipse was not central, the total phase lasted for 51 min. The greatest eclipse occurred at 14:36 UT\cite{nasa}. The Moon\textquoteright s path through the Earth\textquoteright s shadows as well as a map illustrating worldwide visibility of the event is shown in Figure 1. Asia, Australia and part of Pacific had the best visibility. At the observing site, the Moon entered in penumbra before moon-rise and exited before reaching its maximum altitude in the sky.

This paper is arranged as follows: the first section includes motivation for the study and discribes the eclipse parameters. The experimental setup is explained in the section 2 of the paper. In section 3 we discuss our observations of temporal SCR flux variation during the total lunar eclipse and the subsequent control day. Section 4 presents weather conditions and their correlation with SCR flux variation on respective days. Geomagnetic and interplanetary conditions are discussed during both days in section 5. The tidal/gravitational effect is discussed in section 6. Section 7 concludes the paper with discussion based on present investigation.}

\section{Experimental setup}
\label{2}

{
A NaI (Tl) scintillation detector having dimensions of 7.62 cm $\times$ 7.62 cm was used to measure the variation of SCR flux during the eclipse. The detector was shielded by 5 cm thick lead bricks in a rectangular arrangement to minimize the background counts from the Earth and surroundings allowing incoming SCR flux from top. The top view of the detector and lead shielding is shown in Figure 2. The detector signal generated by photo-multiplier tube (PMT) is processed through preamplifier (Pre-Amp), linear amplifier, multi-channel analyzer (MCA) and then stored in a computer. A schematic arrangement of the setup is shown in Figure 3.  The detector was calibrated at regular intervals during the observations using radioactive sources $^{137}$Cs (0.662 MeV), and $^{60}$Co (1.173 MeV, 1.332 MeV and 2.505 MeV (sumpeak)).

\begin{figure}[htp]
\centering
\includegraphics[angle=0,width=100mm]{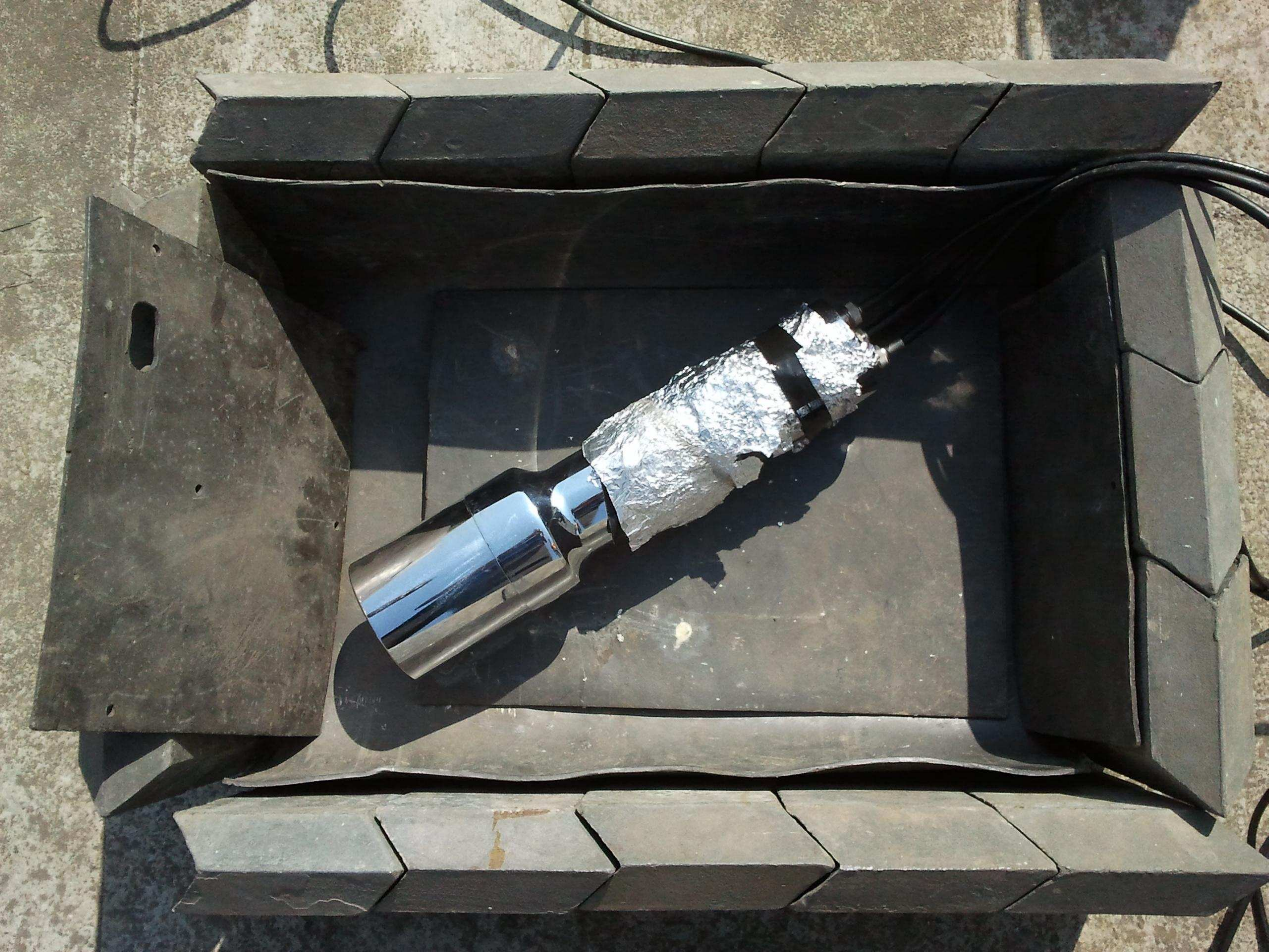}
\caption{Top view of the scintillation detector kept in lead shielding.}
\end{figure}

The SCR flux was recorded with integration time of 10 minutes. To eliminate the possible low energy noise from each spectrum, counts were summed by keeping energy threshold of 200 KeV. To quantify the effect of lead shielding, measurements of SCR flux were carried out with and without lead shielding well before the eclipse started. A significant reduction in the background counts was observed when the detector was shielded as shown in Figure 4. Without shielding, the data showed a background of approximately 87.0 counts per second whereas with shielding a background of approximately 19.8 counts per second was observed. During the eclipse we observed maximum enhancement of 1.6 counts per second which suggests an increase of 8.1 $\%$ over the average of pre and post eclipse data. In the absence of lead shielding, the enhancement would have been 1.8 $\%$ (calculation considered the background count rate of 87 counts per second in case of no shielding) which is not significant. Hence by arranging appropriate shielding one can assure better significance level of variation in SCR flux with respect to the average background
\cite{Jackson01}
\begin{figure}
\centering
\includegraphics[angle=0,width=100mm]{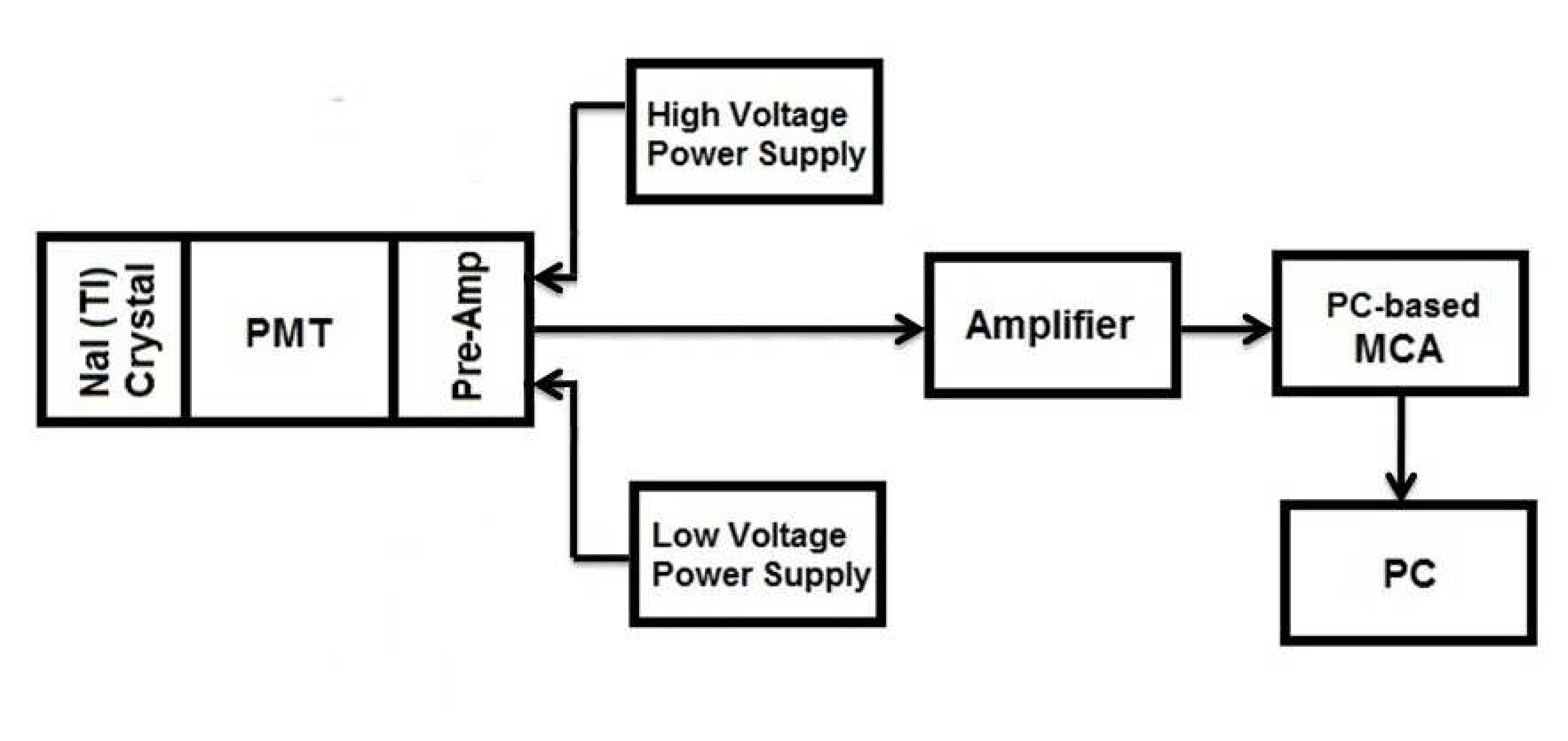}
\caption{The schematic arrangement of the setup used for the SCR flux measurement.}
\end{figure}

\begin{figure}[htp]
\centering
\includegraphics[angle=0,width=100mm]{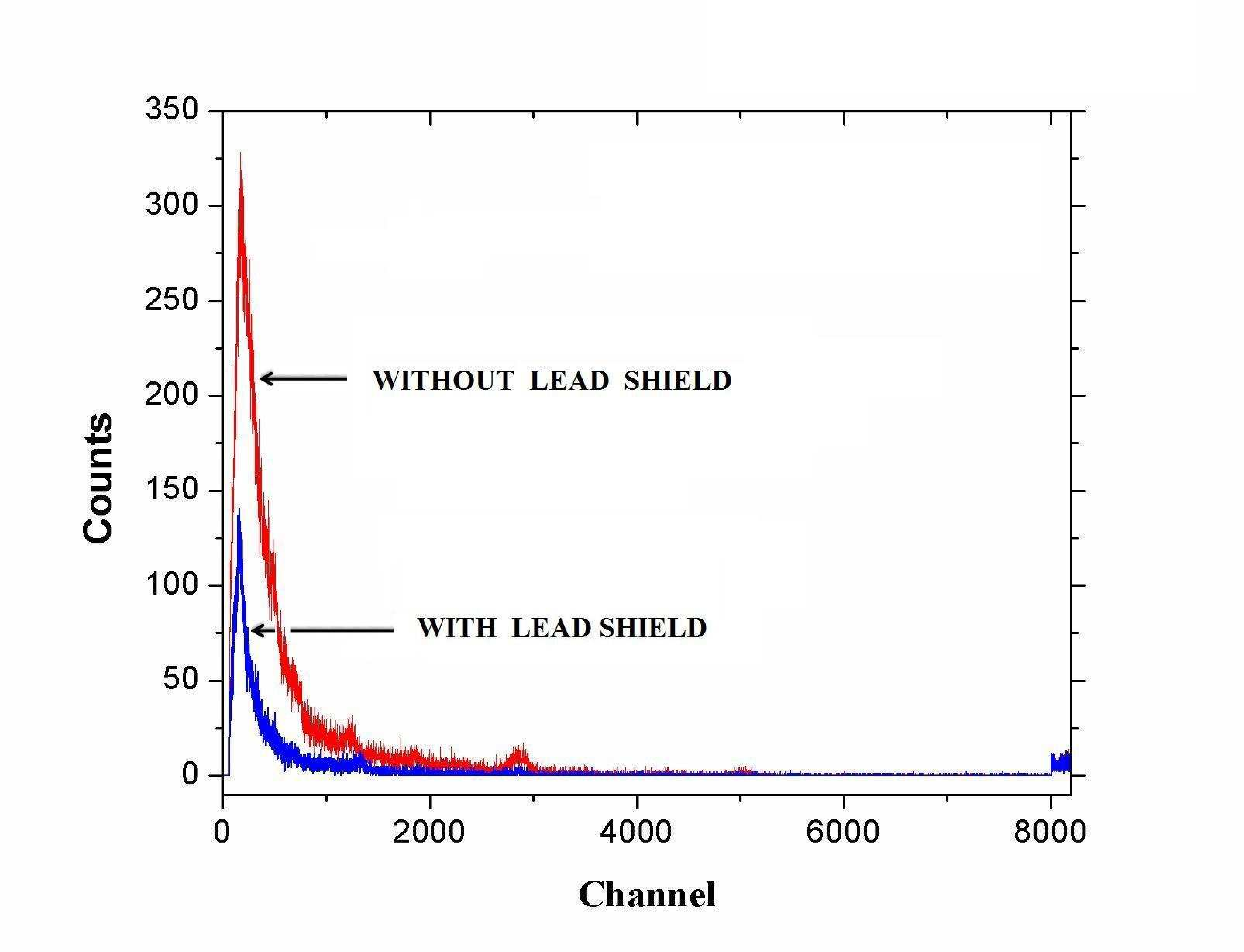}
\caption{Background SCR spectrum with (blue) and without (red) lead shielding.}
\end{figure}

}
\section{Secondary Cosmic Ray flux variation}
\label{3}
{
Temporal variation of SCR flux during the lunar eclipse and the control day is shown in Figure 5. The variation in SCR flux on control day shows increasing trend. The observed enhancement in the SCR flux during the eclipse seems abnormal and unexpected as compared to the general trend of SCR flux on the control day. This enhancement exactly coincides with the lunar entry and exit from the Earth\textquoteright s penumbra at P1 and P4 respectively as shown in the shaded region. This enhancement is prominent and statistically significant. It shows approximately 8.1 $\%$ enhancement in SCR flux during the lunar eclipse when compared to the average of pre- and post-eclipse counts. A double hump structure is clearly seen in SCR flux variation during the eclipse time interval and amplitude of the first hump is higher as compared to the second. It is important to note that the SCR flux before the Moon\textquoteright s entry and exit from the Earth\textquoteright s penumbra converges to the control day SCR flux at corresponding time.

\begin{figure}[htp]
\centering
\includegraphics[angle=0,width=120mm]{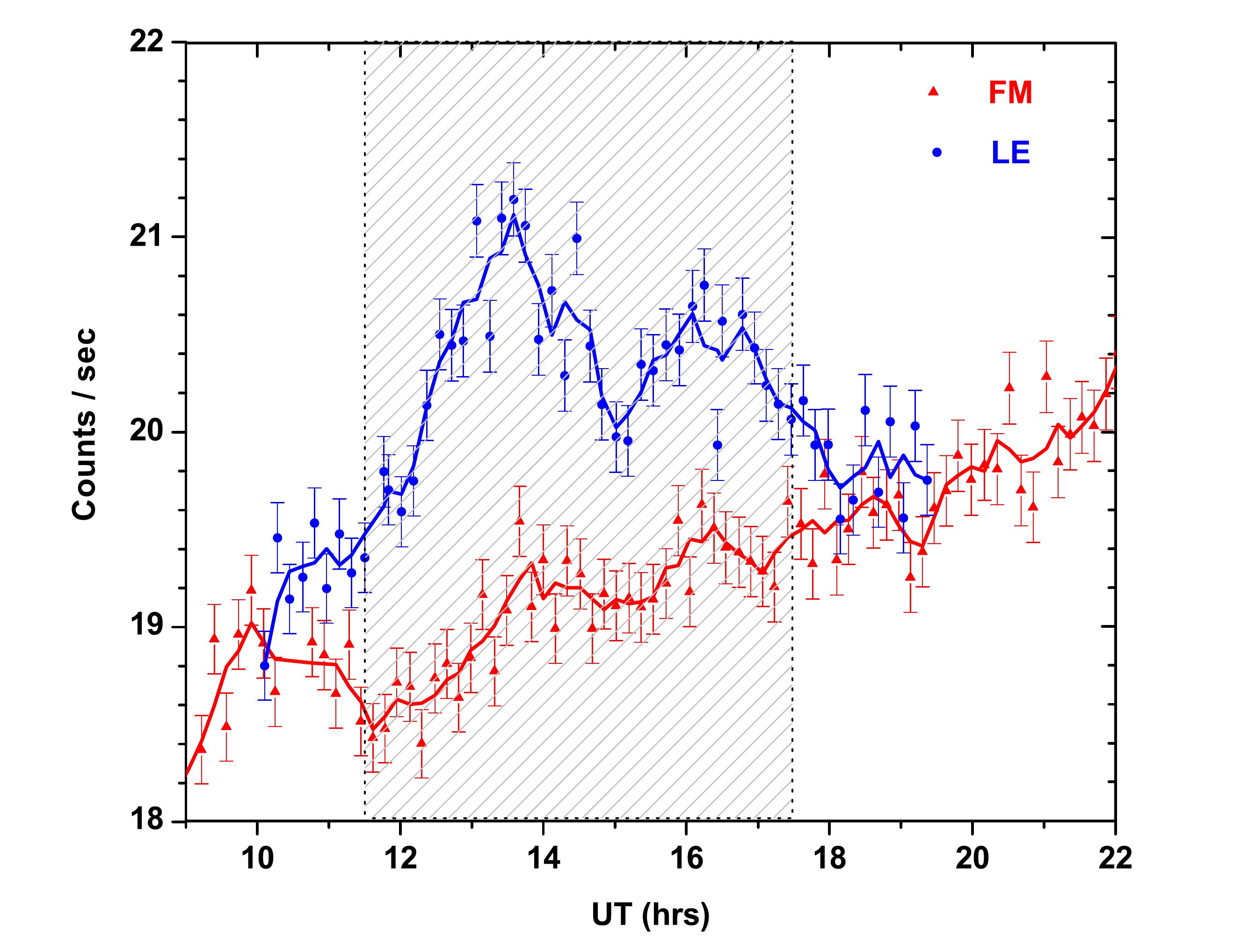}
\caption{SCR flux variation during the lunar eclipse (LE) and control day (FM) recorded by scintillation detector. The shaded region indicates duration of the lunar eclipse.}
\end{figure}

}

\section{Local weather parameters}
\label{4}
{
Ambient temperature and relative humidity were recorded at every 10 minute during the eclipse and control day using digital temperature and humidity sensor. Figure 6 shows temporal variation of relative humidity and temperature during the control and the eclipse days. The shaded region indicates eclipse duration. The diurnal pattern of temperature and relative humidity was clearly seen on both the days. Though weather data is missing at the beginning of the eclipse, it is observed that during the eclipse, temperature steadily decreased from 26.4$^\circ$C to 23.3$^\circ$C and relative humidity increased from 55 $\%$ to 67 $\%$. A similar trend is observed in both parameters on control day during the same time interval. The average temperature/humidity was low/high on control day as compared to the eclipse day. This is due to the control day being in January which is generally cooler than December at the observing site. This assures no abnormal changes in the observed weather parameters during the eclipse.

\begin{figure}[htp]
\centering
\includegraphics[angle=0,width=80mm]{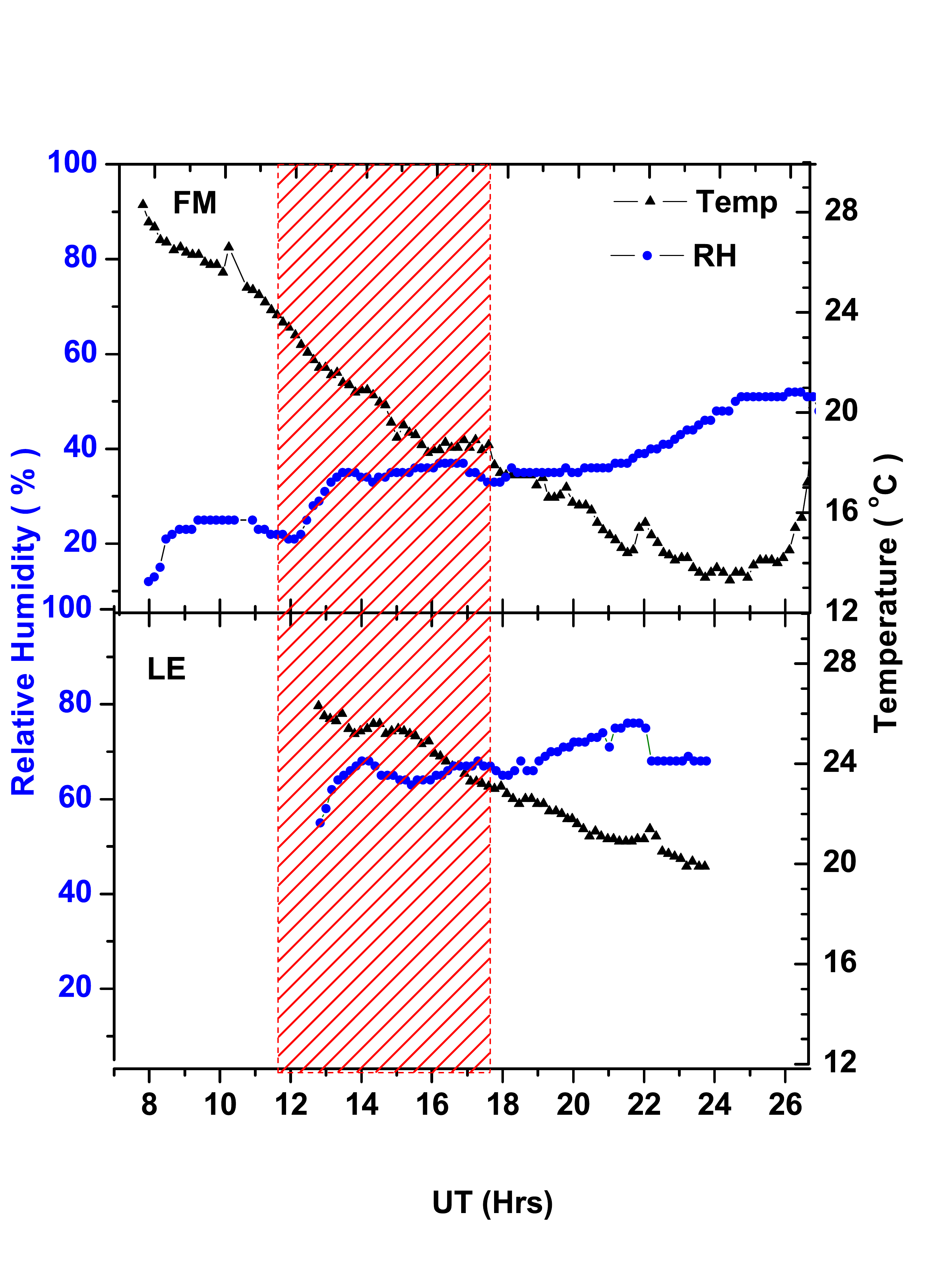}
\caption{The temporal variation of weather parameters (Temperature and relative humidity during control day (FM) and eclipse day (LE). The temperature/relative humidity shows decreasing/increasing trend during eclipse period.  This is due to evening to night transition time at the observation site. Shaded region indicates eclipse duration and corresponding time interval on control day.}
\end{figure}

It is well known that SCR flux gets modulated by weather parameters in which Temperature and humidity play an important role \cite{Lockwood56}. In past studies, anti-correlation of SCR flux with pressure was observed, whereas humidity directly correlated with SCR flux. The effect of temperature is not certain since the Earth\textquoteright s atmosphere is not isothermal, so it is difficult to quantify but negative temperature effect on SCR flux is expected \cite{Olbert53,Clay39}. To investigate the effect of measured weather parameters we carried out the correlation analysis between weather parameters and SCR flux for the eclipse and control day. The estimated Pearson and Spearman correlation coefficients are shown in Table 1.

\begin{table}[ht]
\centering                          % used for centering table
\begin{tabular}{c c c c}            % centered columns (4 columns)
\hline\hline                     %inserts double horizontal lines
     &  & Pearson  & Spearman \\
Case & Correlation between &  Correlation &  Correlation \\% inserts table
     &                     &  coefficient &  coefficient \\%heading

\hline\hline  \\ [0.5ex]                             % inserts single horizontal line
	& Relative Humidity and SCR flux &  0.92 & 0.90 \\ [1ex]           % inserting body of the table
Control day & Temperature and SCR flux & $-$0.78 & $-$0.78 \\ [1ex]
	& Relative Humidity and Temperature & $-$0.84 & $-$0.87 \\ [1ex]
\hline \\ [0.5ex]
	 & Relative Humidity and SCR flux & $-$0.52 & $-$0.50 \\ [1ex]
Lunar Eclipse day & Temperature and SCR flux & 0.80 & 0.80 \\ [1ex]  
 	& Relative Humidity and Temperature & $-$0.73 & $-$0.79 \\ [1ex]     % [1ex] adds vertical space
\hline                              %inserts single line
\end{tabular}
\label{table:nonlin}
\caption{Correlation coefficients of SCR flux and weather parameters.}          % is used to refer this table in the text
\end{table}

As expected high anti-correlation between temperature and relative humidity is observed on both the days. On control day, the temporal variation in SCR flux was strongly correlated with relative humidity and anti-correlated with temperature. Whereas, estimated correlation coefficients(greater than 90 $\%$ confidence level) suggest, SCR flux anti-correlate with relative humidity and well correlate with temperature during the eclipse. The observed correlations of SCR flux with weather parameters during the eclipse are opposite as compared to control day which cannot be explained physically. It is important to note that though two quantities show high correlation, there might not be a cause and effect relationship between the two quantities.  Therefore observed enhancement in SCR flux during the lunar eclipse cannot be explained on the basis of variation in local weather parameters only.

}

\section{Interplanetary and geomagnetic parameters}
\label{5}

{\begin{figure}%[htp]
\centering
\subfigure [(Lunar eclipse)]{\label{fig:le-a}\includegraphics[width=75mm]{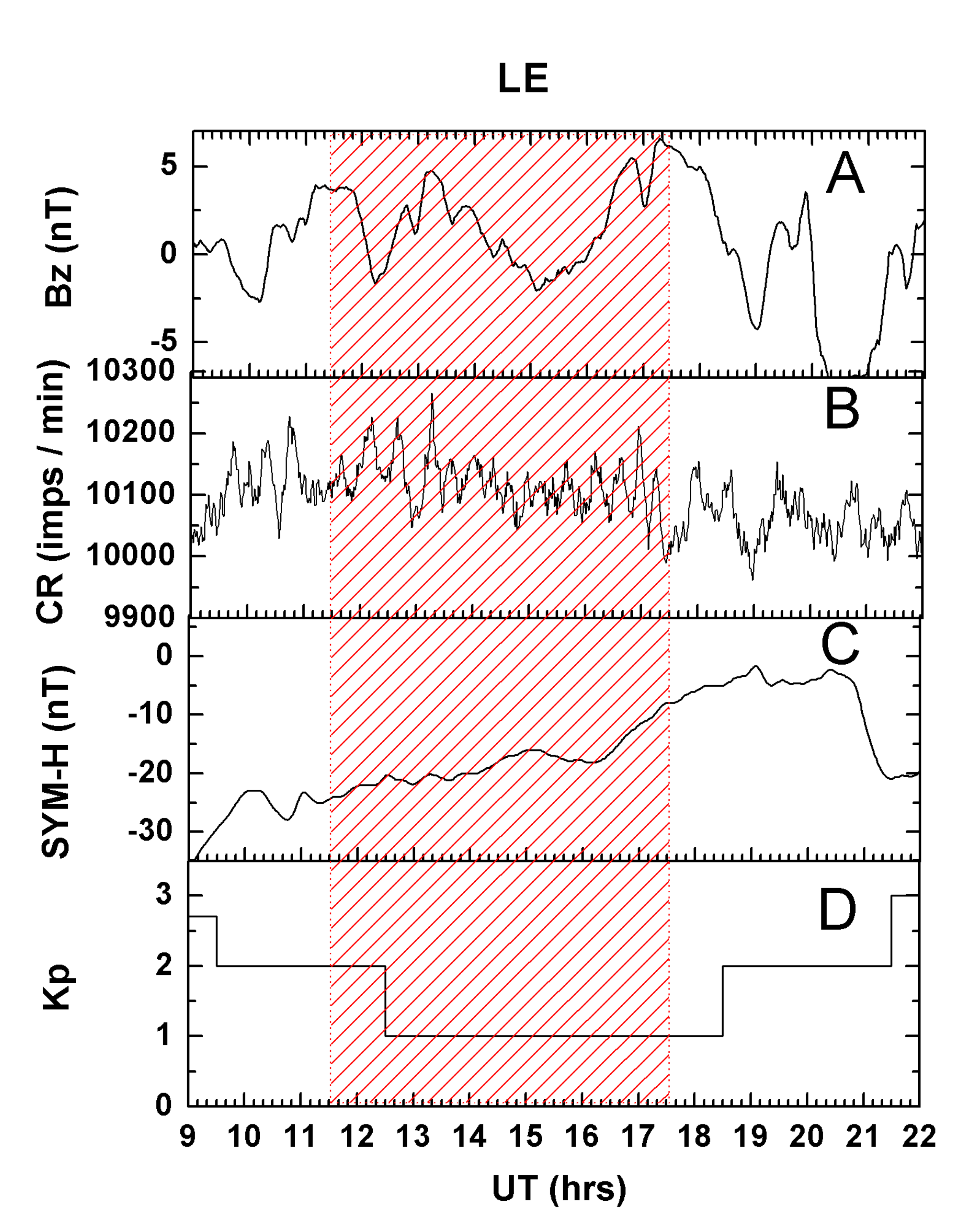}}
\subfigure [(Full-moon)]{\label{fig:le-b}\includegraphics[width=75mm]{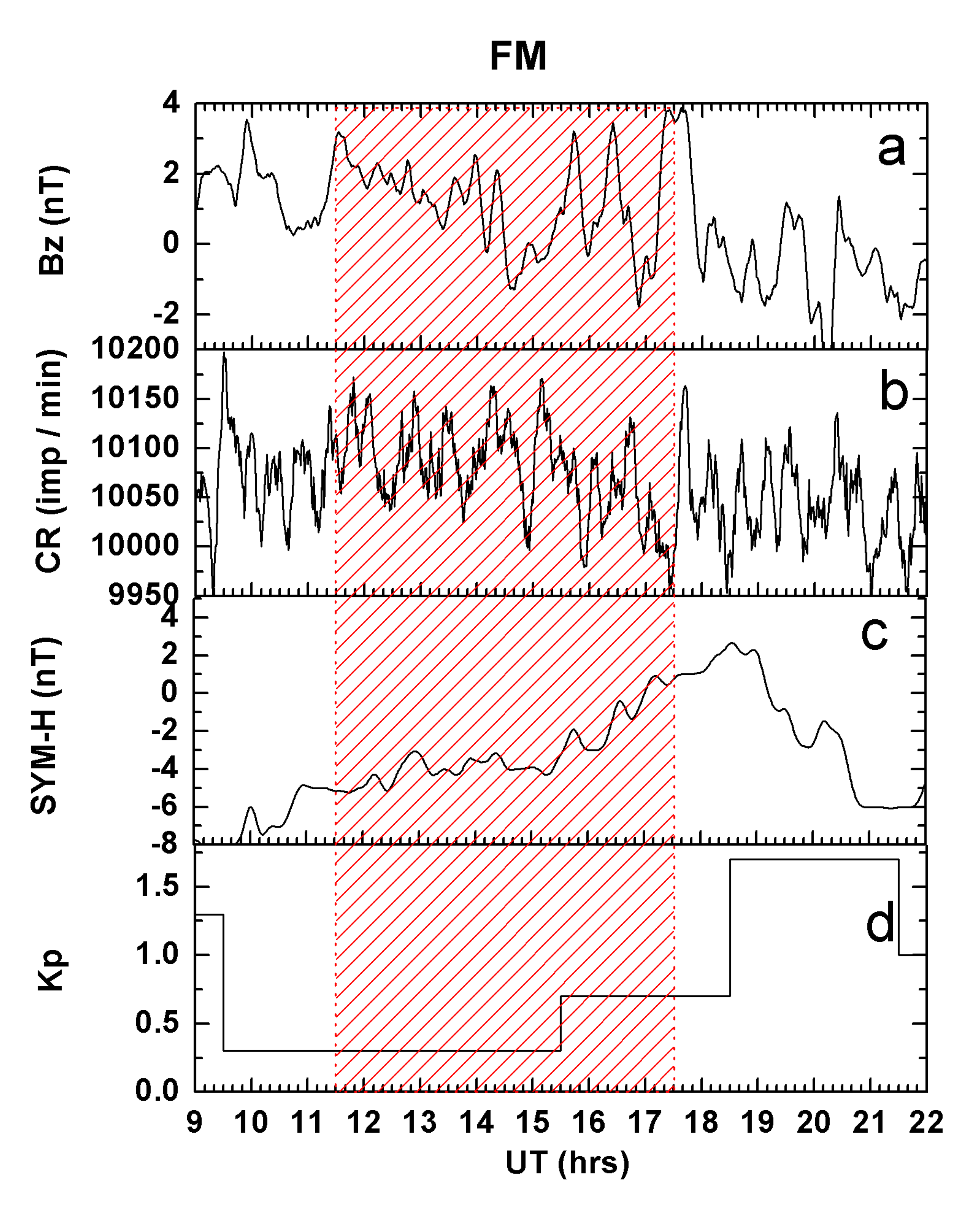}}
\caption{Interplanetary and geomagnetic parameters for eclipse day (left panel) and control day (right panel). [A, a] Interplanetary magnetic field vertical component (Bz). [B, b] Cosmic ray neutron flux (CR). [C, c] Symmetric component of the terrestrial ring current (SYM-H). [D, d] Planetary 3-hr range index (Kp). Shaded region indicates eclipse duration and corresponding time interval on control day.}
\label{fig:le}
\end{figure}

The effects of interplanetary and geomagnetic parameters on SCR flux during the lunar eclipse and control day were studied using interplanetary and geomagnetic indices data from Coordinated Data Analysis Web (CDAWeb) database (http://cdaweb.gsfc.nasa.gov/). The SCR (neutron flux of 10-20 GeV) data obtained from (http://helios.izmiran.rssi.ru/cosray/main.htm) the Moscow Neutron Monitor station. This station did fall in the eclipse visibility region but SCR (neutron flux) shows no systematic variation during the eclipse which can be correlated to SCR flux observed at the site during the eclipse and control days which is seen in Figure 7 (B) and Figure 7 (b) respectively. The vertical component of interplanetary magnetic field (IMF Bz) fluctuated between southward and northward indicating small disturbance in interplanetary medium (Refer Figure 7 (A) and Figure 7 (a)).  The Symmetric-H (SYM-H) \cite{Wanliss06,WDC} index is the strength of symmetric ring current which encircles the Earth in the geomagnetic equatorial belt and generally gets intensified during prolonged southward Bz. Figure (C) and Figure (c) show temporal variation of the SYM-H. The SYM-H shows increasing trend during the eclipse indicating recovery phase of weak geomagnetic storm. Though SYM-H was negative it was less in amplitude ($\leq$30 nT) indicating minor geomagnetic disturbance \cite{Gonzalez94}. Planetary Kp index remained low ($\leq$2) during the eclipse and during the control day assuring a geomagnetic quiet period \cite{Rajaram98}.

\begin{table}[ht]
\centering                          % used for centering table
\begin{tabular}{c c c c}            % centered columns (4 columns)
\hline\hline                     %inserts double horizontal lines
     &  & Pearson  & Spearman \\
Case & Correlation between &  Correlation &  Correlation \\% inserts table
     &                     &  coefficient &  coefficient \\%heading

\hline\hline  \\ [0.5ex]                             % inserts single horizontal line
	& B$_{z}$ and SCR flux &  $-$0.22 & $-$0.3 \\ [1ex]           % inserting body of the table
Control day & SYM-H and SCR flux & 0.72 & 0.80 \\ [1ex]
	& Neutron flux and SCR flux & $-$0.13 & $-$0.14 \\ [1ex]
\hline \\ [0.5ex]
	 & B$_{z}$ and SCR flux & 0.02 & $-$0.02 \\ [1ex]
Lunar Eclipse day & SYM-H and SCR flux & 0.04 & 0.18 \\ [1ex]  
 	& Neutron flux and SCR flux & 0.11 & 0.14 \\ [1ex]     % [1ex] adds vertical space
\hline                              %inserts single line
\end{tabular}
\label{table:correlation}
\caption{Correlation coefficients of SCR flux with Interplanetary and geomagnetic parameters.}          % is used to refer this table in the text
\end{table}

The Moscow Neutron Monitor measures SCR (neutron flux of 10-20 GeV) and the scintillation detector used in the present study measures SCR (0.2 MeV to $\sim$4.5 MeV). Also, the neutron monitor is located at high geomagnetic latitude, whereas the present observations were carried out at low geomagnetic latitude. Due to lower geomagnetic cutoff rigidity at high geomagnetic latitudes one should expect much higher SCR flux over there as compared to low geomagnetic latitudes \cite{Thompson38}. In addition to all these, as explained earlier in section 2, lead shielding was used to minimize the surrounding background radiation. The differences between the two setups and the locations may give rise to the observed discrepancy between the two observations. To investigate the relation between interplanetary and geomagnetic parameters with observed SCR flux, correlation analysis was performed. Results of the analysis are shown in Table 2. There is almost no correlation observed of SCR flux with Bz, SMY-H and neutron flux. On control day, SCR flux shows weak anti-correlation with Bz and Neutron flux where as strong correlation with SYM-H. Thus the absence of any systematic correlation of SCR flux with interplanetary or geomagnetic parameters rules out the interplanetary or geomagnetic origin for observed enhancement.

}

\section{Tidal/Gravitational effect}
\label{6}

{
The Sun and the Moon generate tidal waves in the Earth\textquoteright s magnetosphere, atmosphere and oceans through gravitational interaction. Amplitude of the atmospheric tide increases with altitude so one can expect stronger tides in magnetospheric plasma which can modulate the SCR flux \cite{Mitra51,Appleton39}.  At a distance r (expressed in terms of the Earth\textquoteright s radius) measured from the center of the Earth, the lunar tidal acceleration (a) will be

\begin{equation}
a = GM_M/(R_{ME} - r)^2 - GM_M/(R_{ME}^2)\approx 2rGM_M / R_{ME}^3
\end{equation}

where, $R_{ME}$ is the Moon-Earth distance and $M_M$ is mass of the Moon \cite{Dorman09}. The tidal force is maximum when the Moon reaches the zenith/nadir of the observer so generally high tide is observed when the Moon crosses the zenith/nadir. However, at the observing site the Moon was closer to the horizon at the beginning of the eclipse and the eclipse ended well before the Moon reached its maximum elevation in the sky. Also the observed SCR flux enhancement during the lunar eclipse and decrease during solar eclipse discards the possibility of tidal effect by assuming the relative alignment of the Moon, the Earth and the Sun. Therefore the possibility of tidal effect to explain the enhancement in SCR flux is ruled out.
}

\section{Discussion}
\label{7}
{

In the past, there have been many observations of decrease in SCR flux during solar eclipses. During a solar eclipse, obscuration of the Sun by the Moon brings a rapid change in the intensity of solar radiation causing sudden changes in weather parameters. Decrease in SCR flux has been generally ascribed to the rapidly changing weather parameters during solar eclipses. But, even the recent studies have failed to firmly establish the actual physical mechanism of the phenomenon. It is important to note that the reported observations of SCR flux during solar eclipses show direct correlation between the decrease and temperature, whereas the earlier studies on diurnal variation of SCR and meteorological effects on SCR show anti-correlation with temperature. Therefore, the decrease in SCR flux during a solar eclipse is still an unsolved mystery.

Unlike solar eclipse induced modulation of SCR, lunar eclipse induced SCR modulation was unexpected due to the absence of any rapid change in weather parameters during a lunar eclipse. The cosmic ray shadow effect of the Moon and the Sun (decrease in GCR flux due to the direct blocking of GCR by the Sun or the Moon) has been observed in GCR (in TeV energy regime) by using cosmic ray arrays \cite{Amenomori93a,Amenomori93b}. However, this decrease is always present irrespective of solar or lunar eclipse occurrence. In the present study, the detector was kept at a fixed position irrespective of the Moon\textquoteright s position. Hence, in the present observations by considering the isotropy in GCR flux, no changes in GCR flux and therefore, no change in SCR flux is expected due to the blocking effect. Therefore the possibility of shadow effect of the Moon responsible for enhancement in the SCR flux is ruled out.

In 1967, Anand Rao had studied lunar and solar eclipses by monitoring variation in SCR flux using a GM counter. He had observed enhancements in SCR flux during the lunar eclipses \cite{Ananda67}. Surprisingly his work got unnoticed and after him no one has studied SCR flux variation during a lunar eclipse until the present study. We believe that the present work reports the first observation of SCR flux variation during a lunar eclipse using a scintillation detector. Comparative study of the eclipse and control days indicates that the observed enhancement in SCR flux is unambiguous. We have systematically ruled out the possibility of local weather, interplanetary, geomagnetic parameters and tidal effect which first appeared to be the likely candidates causing the enhancement in SCR flux. This raises a possibility of some unknown parameter/parameters which is/are responsible for the observed enhancement. It is likely that the source causing SCR variation during eclipses lies beyond the Earth\textquoteright s environment and may be is associated with the Moon. At present, the underlying physical phenomenon is unknown and appears to have potential to initiate detailed work. We lay strong emphasis on more comprehensive observations during upcoming lunar eclipses to validate the present observations and investigate the physical mechanism.

}

\section*{Acknowledgments}

{

We are thankful to WDC for Geomagnetism, Kyoto and CDAWeb for making interplanetary and geomagnetic data available. We are thankful to Department of Physics, University of Mumbai, Mumbai for providing experimental resources and facilities. We extend our heartfelt thanks to S. M. Chitre, S. B. Patel and S. K. Tandel of UM-DAE Center for Excellence in Basic Sciences, Mumbai for their constant encouragement and support. We would like to express our gratitude to A. A. Rangwala, D. C. Kothari, A. Misra, C. V. Gurada and R. Srinivasan of Department of Physics. We thank G. Rajaram, G. Vichare and D. Tiwari of Indian Institute of Geomagnetism, Navi Mumbai. We are also thankful to all non-teaching staff of Department of Physics for their cooperation through out the experiment. We are thankful to A. Patwardhan of St. Xavier’s College, Mumbai. We thank Riddhi Kadrekar, Priya Pagare, Katharine Rowlins, Namrata Maladkar, Shraddha Chalke, Satish Vishwakarma, J. More, Bipin Sonawane, S. Kadam and S. Sathian.

}

\end{document}